# COVID-19 Detection Using Modified Xception Transfer Learning Approach from Computed Tomography Images

Kenan Morani [a,1,*], Esra Kaya Ayana [b,2], Devrim Unay [a,3]

[a] Electrical and Electronics Engineering Department, Izmir Democracy University, Izmir, Turkey

[b] Control and Automation Department, Yildiz Technical University, Istanbul, Turkey

[1] kenan.morani@gmail.com*; [2] esrakaya@yildiz.edu.tr; [3] devrim.unay@idu.edu.tr

* corresponding author

ABSTRACT

The significance of efficient and accurate diagnosis amidst the unique challenges posed by the COVID-19 pandemic underscores the urgency for innovative approaches. In response to these challenges, we propose a transfer learning-based approach using a recently annotated Computed Tomography (CT) image database. While many approaches propose an intensive data preproseccing and/or complex model architecture, our method focusses on offering an efficient solution with minimal manual engineering. Specifically, we investigate the suitability of a modified Xception model for COVID-19 detection. The method involves adapting a pre-trained Xception model, incorporating both the architecture and pre-trained weights from ImageNet. The output of the model was designed to take the final diagnosis decisions. The training utilized 128 batch sizes and 224x224 input image dimensions, downsized from standard 512x512. No further da processing was performed on the input data. Evaluation is conducted on the 'COV19-CT-DB' CT image dataset, containing labeled COVID-19 and non-COVID-19 cases. Results reveal the method's superiority in accuracy, precision, recall, and macro F1 score on the validation subset, outperforming VGG-16 transfer model and thus offering enhanced precision with fewer parameters. Furthermore, when compared to alternative methods for the COV19-CT-DB dataset, our approach exceeds the baseline approach and other alternatives on the same dataset. Finally, the adaptability of the modified Xception trasnfer learning-based model to the unique features of the COV19-CT-DB dataset showcases its potential as a robust tool for enhanced COVID-19 diagnosis from CT images.

## 1. Introduction

Within the healthcare realm, the emergence of COVID-19 has been designated as a novel pandemic. The COVID-19 pandemic demands swift and accurate diagnostic methods, with Computed Tomography (CT) imaging emerging as a pivotal tool. CT scan images offer a superior level of detail, providing a comprehensive 360-degree view of the body's structures. Consequently, CT images have also been central in research efforts focused on detecting and diagnosing COVID-19 [1-4]. CT offers rapid imaging, support serial monitoring, detect complications, enable high-risk population screening, contribute to research, and facilitate a needed quick and accurate global response [5-6].

The Polymerase Chain Reaction (PCR) stands as the primary diagnostic test for COVID-19; however, it is marred by drawbacks including supply, machinery, and training expenses. Additionally, accuracy issues with PCR results have been observed [7]. Given these limitations, the exploration of alternative testing methods has become imperative.

Various imaging techniques are employed in COVID-19 diagnosis, including Computed Tomography (CT) and X-ray. Additionally, transfer learning model architectures play a crucial role in the field of medicine [8].

Transfer learning streamlines manual feature engineering, a traditionally labor-intensive process in model development. Instead of starting from scratch, pre-trained models, such as VGG, ResNet, or BERT, are employed. These models come with knowledge acquired from related tasks, allowing for the extraction of lower-level data characteristics like edges and textures. This strategic approach significantly reduces the time and effort required for feature engineering while enhancing efficiency and accuracy in COVID-19 detection from CT images. The choice to incorporate pre-trained weights, particularly from datasets like ImageNet, is rooted in the rich knowledge and robust feature representations they offer. ImageNet's extensive and diverse image collection enables models to comprehend a broad spectrum of visual patterns. By initializing a new model with these pre-trained weights, it inherits an understanding, expediting the learning process. This means the model quickly adapts to the nuances of CT images, capturing COVID-19-specific patterns with greater accuracy. This not only boosts overall performance but also equips the model to address the unique challenges posed by the new dataset, ultimately enhancing its real-world utility in COVID-19 diagnosis.

A different research endeavor utilized the publicly available dataset known as the "COVID-19 Radiology Dataset." This study aimed to assess the capabilities of various pre-trained deep-learning networks in effectively capturing the diverse manifestations of COVID-19 [9]. The comparative analysis revealed that VGG16, MobileNet, DenseNet169, and InceptionV3 outperformed other methods in accurately identifying COVID-19-related manifestations in chest X-ray (CXR) images, showcasing both high sensitivity and accuracy. Nevertheless, VGG16 stood out with better precision.

In [10], a baseline model was introduced for the identification of COVID-19 within an expanded set of CT images called the "COV19-CT-DB" database. This baseline approach utilizes a deep learning technique, specifically a combination of Convolutional Neural Network and Recurrent Neural Network (CNN-RNN network). The model's effectiveness was primarily assessed in terms of the macro F1 score, using both a validation set and a test set from their dataset, which they referred to as the COV-19CT-DB dataset [11-15].

Leveraging the updated version of the COV19-CT-DB dataset, Robert Turnbull carried out another study, as documented in [16]. In this research, a three-dimensional convolutional neural network, referred to as "Cov3d" in the paper, was introduced to detect the presence of COVID-19. This model was constructed upon the 3D ResNet-18 architecture available in the Torchvision library. What's more, during training, a customized loss function was employed. The outcomes demonstrate best performance on the given dataset. However, this performance necessitates complex model architecture and data processing.

In a similar study in [17] on the same dataset, a different strategy was applied, involving a two-stage COVID-19 classification process utilizing BERT features. In the initial stage of BERT feature extraction, a 3D-CNN was deployed to extract internal feature maps from the CNN. Subsequently, a later BERT temporal pooling technique was employed to consolidate the temporal information present within these feature maps, ultimately followed by a classification layer. It's worth noting that the performance of this approach surpassed the performance of the baseline method, yet it possess complexity in model architecture and in data processing.

Regarding methods involving transfer learning, a prior study conducted on the initial release of the COV19-CT-DB dataset in [18]. This study explored and contrasted various transfer learning techniques for COVID-19 classification. The application of "AutoML" techniques, which demand fewer resources, proved to be effective in achieving accurate diagnoses of COVID-19 from 3D volumetric images. Notably, among various methods evaluated, the utilization of the ResNest14 architecture yielded the highest performance in terms of both accuracy and F1-score. The method, although utilizes a pretrain architecture with little processing, still maintains relatively high number of training parameters due to the chosen model architecture.

Numerous approaches outlined in the literature have offered solutions for COVID-19 diagnosis and detection, with transfer learning being a central component in many of these solutions. However, these methods will usually be complex, utilizing image processing techniques or heavy models with a lot of trainable parameters. This paper introduces a transfer learning approach based on Xception to harness the capabilities of this neural network architecture while maintaining simplicity and speed in the methodology. Minimal image processing was utilized, and fewer number of parameters were used. With that we put forward a method that gives an efficient performance, while also offering an easily understandable algorithm.

Outlined below are the primary contributions of this paper:

- Introduction of an adapted transfer learning model for COVID-19 detection with Xception model as a baseline.
- Conduction of a performance assessment, comparing our model against other transfer learning approaches on a both validation and a test set of unseen images.
- Demonstration of the efficacy of our proposed solution in enabling data-driven decisions, achieving comparative performance, which exceeds the baseline and other approaches on challenging and big CT image database.

## 2. Material and Method

In our methodology, the adopted/modified architecture to the transfer learning model involves a global average pooling, a dense layer incorporating 128 filters with Rectified Linear Units (ReLU) activation, followed by batch normalization, a 0.2 dropout, and culminating in a dense layer using a sigmoid activation function. Fig. 1 Shows the model architecture output.

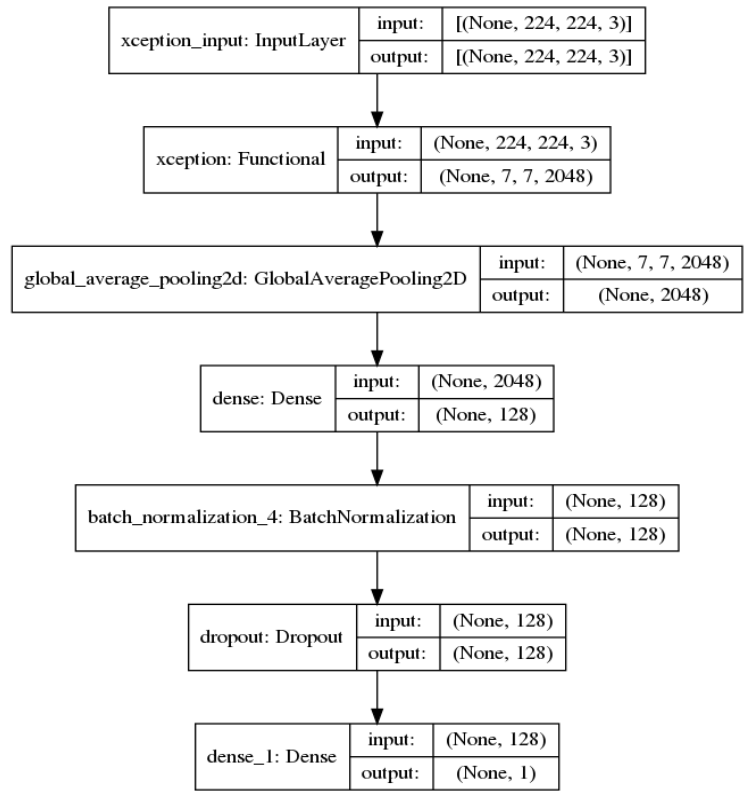

**Fig 1** Adapted Xception model architecture

The final layer's output represents the class probability of being a non-COVID-case slice. This class probability is then compared against a predefined threshold, determining the slice's classification as COVID or non-COVID. These individual slice-level determinations collectively lead to patient-level diagnoses, as explained in the latter sections of the paper.

Different class probability thresholds were tried and compared for the best performance. The results of using different class probability thresholds were compared both on the validation set and on the test set in the results section.

Several class probability thresholds were explored to optimize performance, and their effects were evaluated on both validation and test sets. The model's architectural composition results in a total of 21,124,393 parameters, with 262,657 being trainable.

In our training process, the transfer-learning models were employed with a 3-channeled input, harnessing pre-trained weights from the ImageNet model. Notably, we rendered the model's weights non-trainable during our work's training phase. To enhance the training dynamics, callback mechanisms were implemented. Specifically, the "ReduceLROnPlateau" callback was utilized [19-22]. The "ReduceLROnPlateau" mechanism is a dynamic learning rate adjustment strategy commonly used during the training of deep learning models. Its purpose is to enhance the convergence process by adapting the learning rate in response to the observed performance of the model during training. In our case, the "ReduceLROnPlateau" callback was particularly useful due to its relevance in optimizing the performance of our COVID-19 detection model. This mechanism operates by continuously monitoring a specified metric, which in our implementation was the validation loss. If the validation loss does not demonstrate improvement over a predefined number of training epochs (in this case, a patient of 2 epochs), the learning rate is adjusted downward by a certain factor (often referred to as the "factor" parameter). This mechanism aligns well with our objective of training a COVID-19 detection model. The complex and multifaceted nature of COVID-19-related patterns in CT images could potentially lead to slower convergence or plateaus in the validation loss. By using the "ReduceLROnPlateau" callback, we adaptively addressed this challenge. When the validation loss stagnates, the learning rate reduction acts as a corrective step, allowing the model to navigate through regions of loss landscapes where progress may be hindered.

For model compilation, we employed the Keras platform, utilizing the "Adam" optimizer initialized with a learning rate of 0.001. The loss function was set as "binary cross entropy." "Adam" stands out as an adaptive learning rate optimization algorithm that dynamically adjusts learning rates for each parameter during training. This adaptability allows the model to converge more efficiently and effectively navigate complex loss landscapes, potentially accelerating training and improving convergence speed.

Furthermore, the selection of "binary cross entropy" as the loss function aligns well with the nature of our COVID-19 detection task. Binary cross entropy is a suitable choice for binary classification tasks like ours, where we distinguish between COVID-19 and non-COVID-19 cases. It quantifies the dissimilarity between predicted and actual class labels, facilitating the model's efforts to accurately distinguish between the two classes.

Our model was trained across 13 epochs, a determination that emerged from rigorous experimentation. During these trials, it was observed that further increasing the epoch count resulted in only marginal improvements in validation loss over a prolonged timeframe.

Training the CNN model, with a batch size of 128, across the 13 epochs necessitated approximately 7 days of computation. This was facilitated on a workstation operating a GNU/Linux system, equipped with 64GiB of system memory and powered by an Intel(R) Xeon(R) W-2223 CPU @ 3.60GHz processor. These specifications offer insights into the computational resources and time investment involved in achieving our model's refined performance for COVID-19 detection.

The input images, resized to 224x224 dimensions, are converted from the original 512x512 grayscale images, with no further manipulation applied to the initial slices.

## 2.1 Xception Based Transfer Leanirng Model

The proposed transfer learning approach is rooted in a modified Xception model, renowned for its 36 convolutional layers that establish the foundation for feature extraction within the network. The core of this model is comprised of the Xception architecture, followed by an adapted output structure.

The selection of the Xception architecture stemmed from its compelling attributes, including its adaptability to complex image datasets and efficient computation. The model's depth-wise separable convolutions and shortcuts between convolution blocks contribute to its lightweight yet powerful nature, making it particularly suitable for our COVID-19 detection task. Furthermore, the decision to borrow pre-trained weights from the ImageNet dataset is underpinned by the principle of transfer learning, which leverages knowledge learned from one task to improve performance on another. ImageNet, with its extensive and diverse collection of images spanning various categories, furnishes the Xception model with foundational features that generalize well across different visual recognition tasks. This approach significantly accelerates the learning process during fine-tuning of our COVID-19 CT image dataset. By initializing our model with these pre-trained weights, we harness a rich set of learned features, saving substantial time and resources that would otherwise be expended on training from scratch. The transfer of these features to our dataset, coupled with the model's intrinsic architecture, enhances its capability to discern intricate patterns in COVID-19-related lung scans. This, in turn, augments the model's diagnostic performance and positions it favorably for accurate and efficient COVID-19 detection [23 - 25].

The Xception classification accuracy using ImageNet data surpasses that of alternative transfer learning options. Comparative data, as shown in Table 1, further corroborates the efficacy of the Xception model in terms of parameter count and accuracy, solidified by its unique convolutional architecture and depth-wise separable convolutions.

**Table 1** Xception vs. other popular pre-trained models

| Model | No. of parameters | Top-1/Top-5 accuracy |
|---|---|---|
| Xception | 22.85 million | 0.790 / 0.945 |
| Inception | 23.62 million | 0.782 / 0.941 |
| VGG | 128 million | 0.715 / 0.901 |
| ResNet | 23 million | 0.770 / 0.933 |

By going through the comparisons as presented in the table above, it can be said that the Xception model outperforms its peers not only in terms of having fewer parameters to train but also in terms of classification accuracy. This justifies our choice of modified Xception transfer learning for COVID-19 detection in this paper.

To further evaluate our Xception based transfer learning model results, we compare it to a VGG16 based model with the same settings. With that in mind, VGG-16 pre-trained architecture and weights were borrowed. To make the final medical diagnosis, an output architecture similar to the one used for the modified Xception approach was implemented.

To evaluate and benchmark our selected transfer learning model against a good alternative ImageNet-pretrained model on the COV19-CT-DB dataset, we incorporated the VGG16 model architecture. Notably, the VGG16 model's ImageNet-trained weights carry a substantial size, totaling 528 MB. This considerable size is often indicative of the model's complexity, which, in turn, tends to correspond with robust performance.

The distinctive character of the VGG16 model architecture predominantly emanates from a deliberate design choice. Instead of proliferating hyperparameters, the emphasis was placed on the utilization of convolutional layers featuring 3x3 filters, each operating with a stride of 1 and consistently employing the 'same' padding. Additionally, a recurring architectural element was the utilization of max-pooling layers equipped with 2x2 filters and a stride of 2. These architectural decisions contribute to the unique characteristics of the VGG16 model and underlie its efficacy in various computer vision tasks [26-27].

## 2.2 The Dataset

The dataset employed in this study is an expansion of the COV19-CT-DB. This dataset plays a pivotal role in our study, offering a comprehensive collection of CT scans that are instrumental for COVID-19 detection. This dataset comprises a substantial number of CT scans, consisting of 1,650 cases of COVID-19 and 6,100 non-COVID-19 instances. This balanced distribution allows for a robust evaluation of your proposed method's performance across different classes.

What sets the 'COV19-CT-DB' dataset apart is not just its size, but also its diversity in terms of both the number of cases and the variability within COVID-19 manifestations. The dataset's annotated CT scans have been meticulously labeled by a panel of experts, each with over 20 years of experience. This ensures the accuracy and reliability of the labels, which is crucial for building and evaluating machine learning models.

The dataset's diversity, which spans a range of COVID-19 and non-COVID-19 cases, presents unique challenges and opportunities. COVID-19 is known to exhibit a spectrum of manifestations, from mild to severe, and capturing this variability is vital for the development of an effective detection model. The 'COV19-CT-DB' dataset's inclusion of cases with varying levels of lung involvement and diverse clinical presentations mirrors the real-world complexity of COVID-19 cases.

Given its richly labeled nature, extensive size, and diversity, the 'COV19-CT-DB' dataset provides an ideal foundation for evaluating the efficacy of your proposed method. Its suitability stems from its ability to rigorously assess your model's performance on different types of cases, ensuring that the method is not only accurate but also robust in identifying COVID-19 instances within varying clinical contexts.

Each CT scan comprises a variable number of slices, ranging from 50 to 700. Access to this dataset is made possible via the "ECCV 2022: 2nd COV19D Competition". The distribution of COVID-19 and non-COVID-19 cases for our study is shown in Table 2.

**Table 2** Distribution of cases in the COV19-CT Database

| Annotation | Training Data | Validation Data | Test Data |
|---|---|---|---|
| COVID-19 CT cases | 882 | 215 | 5281 (Labels are not provided by the organizers) |
| Non-COVID CT cases | 1110 | 269 used out of the original 289 | |

### 2.3 Patient Level Performance

At the patient level, we systematically experimented with and compared various class probability thresholds, leveraging class prediction probabilities to optimize diagnostic accuracy. These thresholds were established based on the probability of class 1 (Non-COVID) prediction. In essence, if the model's output probability for class 1 surpassed the designated threshold, the slice was classified as Non-COVID; otherwise, it was categorized as COVID. In the scenario where the count of COVID-labeled slices equaled the count of non-COVID slices within any CT volume, the ultimate patient-level determination was that the patient was non-COVID. This slice-level decision-making process can be succinctly expressed as follows:

Following the acquisition of slice-level predictions, a patient's diagnosis hinges on the presence or absence of COVID-19-labeled slices within their CT data. Specifically, if the patient's CT dataset comprises a greater number of predicted non-COVID slices compared to predicted COVID slices, the patient is diagnosed as non-COVID, applying a majority voting method. Conversely, if the count of predicted COVID slices surpasses that of non-COVID slices, the patient is diagnosed with COVID.

The clinical significance of the patient-level diagnostic approach we've outlined can be elucidated as follows: Imagine a scenario where a patient exhibits lung damage attributable to COVID-19 in approximately 40% of their CT slices. Consequently, our neural network classifies approximately 40% of these slices as COVID and the remainder as non-COVID. As a result, the ultimate diagnosis aligns with the majority voting principle, and the patient is classified as non-COVID. This approach reflects the real-world clinical interpretation when considering the overall health status of the patient.

Although even a minor anomaly detected in a single slice may be associated with a disease, our speculation within the context of COVID is that a substantial degree of involvement is typically required to inform a diagnostic decision. It's worth noting that our deep learning model exhibits a remarkable level of sensitivity, capable of discerning even the slightest abnormalities in the slices.

Expanding on our earlier observations, it's important to highlight that we explored an alternative approach, which we referred to as the "all-or-nothing approach." In this scenario, a COVID diagnosis is triggered if even a single slice is predicted as COVID-19. However, this approach yielded less accurate results, as detailed in the results section, corroborating the insights we discussed previously.

### 2.4 Performance Evaluation

The proposed model was evaluated via the COV19-CT-DB database using accuracy, macro F1 score, and confidence interval.

The accuracy is calculated as in (1).

$$Accuracy = \frac{True\ Positives + True\ Negatives}{True\ Positives + False\ Positives + True\ Negatives + False\ Negatives} \quad (1)$$

Where positive and negative cases refer to COVID and non-COVID cases.

The macro F1 score was calculated after averaging precision and recall matrices as in (2).

$$Macro\ F1 = \frac{2 \times average\ precision \times average\ recall}{average\ precision + average\ recall} \quad (2)$$

Furthermore, to report the confidence intervals of the results obtained, the Binomial proportion confidence intervals for the macro F1 score are used. The confidence intervals were used to check the range variance of the reported results. The residuals of the interval can be calculated as in (3) [28-29].

$$Radius\ of\ Interval = z \times \sqrt{\frac{macro\ F1 \times (1 - macro\ F1)}{n}} \quad (3)$$

where z is the number of standard deviations from the Gaussian distribution and n is the number of samples.

## 3 Results and Discussion

The results of our methodology are discussed on the validation set and unseen set of images, test set.

### 3.1. Results on the Validation set

In all three approaches, we tackled a binary classification task focused on the detection of COVID-19 from CT images. Our findings indicate that the modified Xception model achieved an average validation accuracy of 0.747. Fig. 2 illustrates the progression of training and validation accuracy, while Fig. 3 visualizes the development of training and test precision and recall metrics for this model.

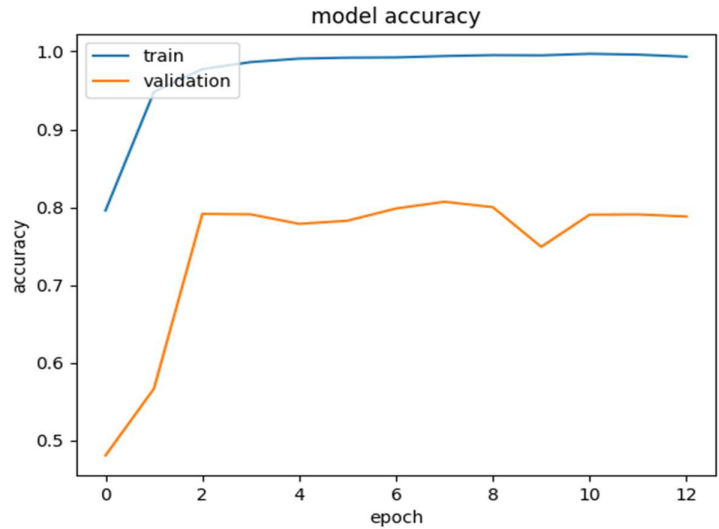

**Fig 2** Evolution of training and validation accuracy

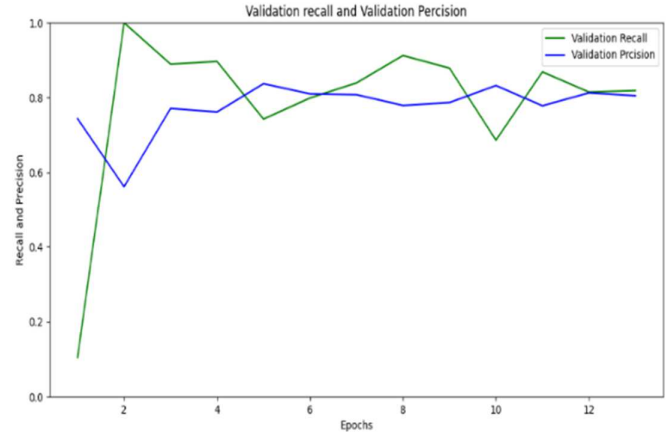

**Fig 3** Evolution of validation recall and precision

Table 3 further shows the training performance for different metrics.

**Table 3** Performance results of the training

| Performance metric | Score |
|---|---|
| Average training accuracy | 0.973 |
| Average recall | 0.788 |
| Average precision | 0.776 |
| Macro F1 score | 0.782 |

To calculate the confidence interval for the resulting macro F1 score, equation 3 was used. In the equation, z is taken as z=1.96 for a significance level of 95%. By that we can obtain the confidence interval for the macro F1 score, keeping in mind that The number of samples (slices) in the validation set is 106,378, to be approximately 0.78. With that, the macro f1 score can be said to be 0.78232 ± 0.00024.

Using the above-mentioned method, predictions were made through different class probability thresholds at the patient level using the majority voting method for each CT scan. Fig. 4 shows performance results on the validation set for four different thresholds. The comparison was made in terms of the validation accuracy and the macro F1 score.

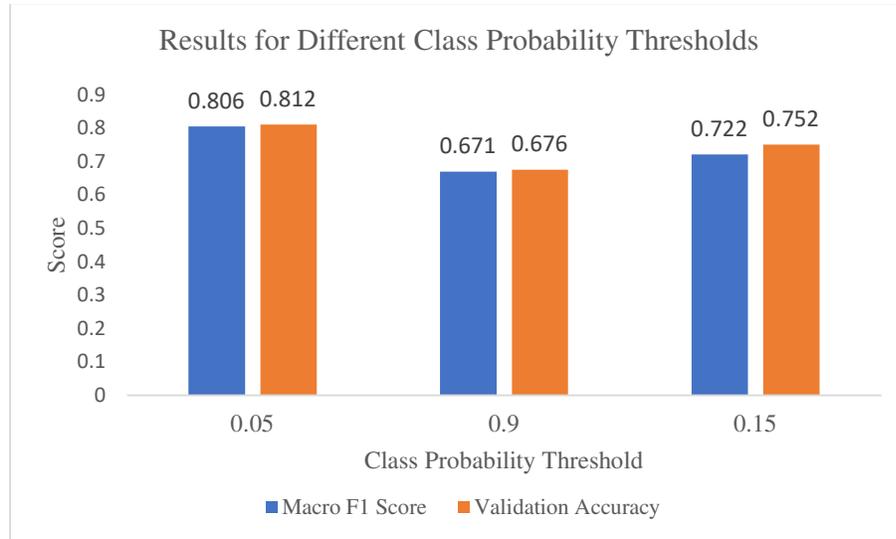

**Fig 4** Model performance against different class probability thresholds on the validation set

The findings indicate that, among the three suggested class probability thresholds, the 0.5 threshold level gives the best performance. This holds when considering both validation accuracy and validation macro F1 score. Consequently, our proposed approach outperforams the baseline model approach, as reported in [10], in terms of macro F1 score, achieving a score of 0.777 on the validation set.

Further evaluate the Xception model by comparing it to other transfer learning methods, the VGG16 based method was used as described in the 'Material and Method' section. The results of the comparison on the validation set can be seen in Table 4. The comparison is in terms of macro F1 score for different class probability thresholds.

**Table 4** Macro F1 score performance comparison (Xception vs. VGG16) on the validation set

| Proposed class probability threshold | VGG16 macro F1 score | Xception macro F1 score |
|---|---|---|
| 0.5 | 0.756 | 0.806 |
| 0.9 | 0.406 | 0.671 |
| 0.15 | 0.505 | 0.722 |

The results establish the better performance utilized by using the modified Xception model over VGG16, notably evident in terms of the macro F1 score. This distinction is further emphasized when considering class probability thresholds of both 0.5 and 0.15. All in all, the results on the validation set of macro F1 score validate our choice of the transfer learning model as well as it is efficiency in COVID-19 Detection on the given dataset.

The proposed Xception model misclassifies, especially for the topmost and bottommost slices within the CT volume. It's worth noting that these extreme slices correspond to anatomical regions where COVID-19 involvement is typically absent. Consequently, these slices are considered less representative of the disease diagnosis process. For a visual understanding, Fig. 5 showcases exemplary slices accurately classified by our proposed model, while Fig. 6 illustrates exemplary slices that were incorrectly classified, prominently featuring the aforementioned extreme slices. It's important to note that these slices were sampled from the validation partition.

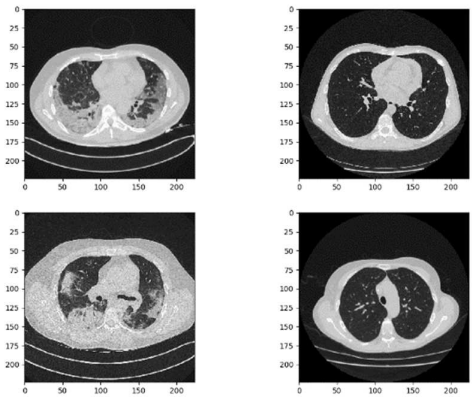

**Fig 5** Examples of correctly classified slices from COVID (right) and Non-COVID (left) cases

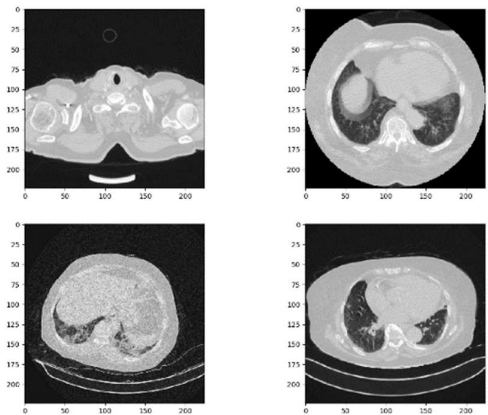

**Fig 6** Examples of misclassified slices from COVID (right) and Non-COVID (left) cases

To delve deeper into the comparative analysis, we extended our evaluation to the test partition, which comprised previously unseen images. In the forthcoming sections, we present a comprehensive examination of these models, offering detailed insights into their performance.

### 3.2. Results of the Test Partition

To further validate our findings, we subjected the methodology to evaluation on the test partition of the COV19-CT-DB database. Within this assessment, we employed various class probability thresholds as part of the methodology. Our proposed approach demonstrated high macro F1 scores when utilizing class probability thresholds of 0.15 and 0.5, respectively.

In addition, we conducted a comparative analysis by pitting our methodology against the VGG16 model on the test partition. The results, summarized in Table 5, consistently show a better performance using the pre-trained Xception model architecture used in our paper, as evidenced by macro F1 scores. These findings reaffirm the robust performance of our proposed approach on previously unseen data.

Table 5 Macro F1 score performance comparison (Xception vs. VGG16) on the test set

| Proposed class probability threshold | VGG16 macro F1 score | Xception macro F1 score |
|---|---|---|
| 0.5 | 0.789 | 0.809 |
| 0.15 | 0.613 | 0.818 |

Further, the f1 score for COVID and the f1 score for non-COVID for our proposed Xception methodology came 0.666 and 0.952 at 0.5 threshold and 0.968 and 0.667 at 0.15 threshold, respectively.

Table 6 Average macro F1 score results from the comparison of validation and test partitions

| The Method | Validation set | Test set |
|---|---|---|
| Cov3d [16] | 0.947 | 0.878 |
| BERT method [17] | 0.916 | 0.808 |
| Base Line [10] | 0.77 | 0.67 |
| **Proposed method (best performance)** | **0.806** | **0.818** |

The results on the test set of unseen images, further validate our Xception transfer-learning based model architecture and proof it efficiency for the task of COVID-19 detection.

The achievement of higher accuracy and macro F1 score with the modified Xception model is an important advancement in COVID-19 detection. This outcome shows the effectiveness of our approach in accurately identifying COVID-19 cases from CT images. Enhancing accuracy is particularly crucial in minimizing false positives and negatives, which are critical concerns in a clinical setting. The higher F1 score demonstrates a balanced performance between precision and recall, essential for reliable diagnoses.

The practical implications of this achievement are profound. In the clinical context, accurate and reliable COVID-19 detection is vital for guiding patient management, isolation, and treatment decisions. Healthcare professionals can use these results to expedite patient triage, allocate resources efficiently, and provide timely care to those in need. As our results were established on a big, diverse and unbalanced data, our results are suited for healthcare settings where data is mostly unbalanced. This dataset can further the scientific community's understanding of COVID-19, leading to improved strategies for disease management.

It is important to recognize limitations in our methodology, including the adaptability of the modified Xception model to diverse datasets and clinical scenarios. To address these, future research directions include exploring the model's versatility across different image modalities, such as X-ray modalities, and investigating alternative output modification strategies, such as employing Random Forest as an output classifier.

## 4 Conclusion

In conclusion, we have proposed a transfer learning-based approach for COVID-19 detection via CT images. The proposed method briefly processes the images in terms of size. We use the Xception model architecture and weights pre-trained on the popular ImageNet. Modification of the network's output was made to take the final diagnosis decisions. The model implemented along with a decreasing learning rate and hyperparameter turning proves effective on a big and challenging dataset of CT images. We, therefore, provide a presentation of our tailored model method, demonstrating our model's superiority over other transfer learning models on both validation and test datasets and the effective enablement of data-driven decisions, resulting in efficient diagnostic performance on a challenging and big CT image dataset for COVID-19 detection. This positions our solution as suitable for clinical applications, particularly where less hand-engineered systems are preferred.

The primary findings and contributions of this study underscore the vital role of transfer learning in COVID-19 detection, particularly when applied to the domain of CT images. This approach not only bolsters diagnostic precision but also holds substantial potential for real-world applications, promising to expedite early detection efforts and mitigate the virus's spread.

It is important to clarify the rationale behind our methodology, aligning it with the principles of the scientific method. Our approach was meticulously crafted to ensure the validity and reliability of our findings. We leveraged transfer learning with the Xception model because it has demonstrated exceptional performance in diverse image recognition tasks, aligning with the scientific method's emphasis on leveraging established knowledge. Pre-training on ImageNet, a vast and varied dataset, serves as a form of hypothesis testing, allowing us to start with a well-informed starting point, analogous to formulating an educated hypothesis in the scientific method.

In summary, our study emphasizes the significance of transfer learning approaches as a potent tool in the fight against COVID-19, especially in the realm of CT image-based diagnosis. Its broader implications underscore its potential to revolutionize diagnostic processes and contribute significantly to public health efforts, even as we continue to refine our approach.

## Acknowledgment

The authors acknowledge the medical staff who rigorously annotated the dataset used for this study and others who shared the dataset with the "IDU-CVLab team".

## Declarations

**Author contribution.** The first author did the research and applied the methodology under the supervision of both the second and the third authors.
**Funding statement.** No funding was provided for this study.
**Conflict of interest.** The authors declare no conflict of interest.
**Additional information.** The code related to this study can be found on Github at https://github.com/IDU-CVLab/COV19D_2nd.

## Data and Software Availability Statements

Data are provided and should be ordered from the corresponding entity as stated in the 'Material and Methodology' section, 'Dataset' subsection.

## References

[1] Duong, Linh T., et al. "Automatic detection of Covid-19 from chest X-ray and lung computed tomography images using deep neural networks and transfer learning." Applied Soft Computing 132 (2023): 109851.

[2] Ağralı, Mahmut, et al. "DeepChestNet: Artificial intelligence approach for COVID-19 detection on computed tomography images." International Journal of Imaging Systems and Technology 33.3 (2023): 776-788.

[3] Asif, Sohaib, et al. "A deep learning-based framework for detecting COVID-19 patients using chest X-rays." Multimedia Systems 28.4 (2022): 1495-1513.

[4] Li, Guang, et al. "COVID-19 detection based on self-supervised transfer learning using chest X-ray images." International Journal of Computer Assisted Radiology and Surgery 18.4 (2023): 715-722.

[5] Kathamuthu, Nirmala Devi, et al. "A deep transfer learning-based convolution neural network model for COVID-19 detection using computed tomography scan images for medical applications." Advances in Engineering Software 175 (2023): 103317.

[6] Msemburi, William, et al. "The WHO estimates of excess mortality associated with the COVID-19 pandemic." Nature 613.7942 (2023): 130-137.

[7] Alyafei, Khalid, et al. "A comprehensive review of COVID-19 detection techniques: From laboratory systems to wearable devices." Computers in Biology and Medicine (2022): 106070.

[8] Kora, Padmavathi, et al. "Transfer learning techniques for medical image analysis: A review." Biocybernetics and Biomedical Engineering 42.1 (2022): 79-107.

[9] Asif, Sohaib, et al. "Detection of COVID-19 from chest X-ray images: Boosting the performance with convolutional neural network and transfer learning." Expert Systems 40.1 (2023): e13099.

[10] Kollias, D., Arsenos, A. and Kollias, S., 2022. Ai-mia: Covid-19 detection & severity analysis through medical imaging. arXiv preprint arXiv:2206.04732.

[11] Kollias, D., Arsenos, A., Soukissian, L. and Kollias, S., 2021. Mia-cov19d: Covid-19 detection through 3-d chest ct image analysis. In Proceedings of the IEEE/CVF International Conference on Computer Vision (pp. 537-544).

[12] Kollias, D., Bouas, N., Vlaxos, Y., Brillakis, V., Seferis, M., Kollia, I., Sukissian, L., Wingate, J. and Kollias, S., 2020. Deep transparent prediction through latent representation analysis. arXiv preprint arXiv:2009.07044.

[13] Kollias, D., Vlaxos, Y., Seferis, M., Kollia, I., Sukissian, L., Wingate, J. and Kollias, S., 2020, September. Transparent adaptation in deep medical image diagnosis. In International Workshop on the Foundations of Trustworthy AI Integrating Learning, Optimization and Reasoning (pp. 251-267). Springer, Cham.

[14] Kollias, D., Tagaris, A., Stafylopatis, A., Kollias, S. and Tagaris, G., 2018. Deep neural architectures for prediction in healthcare. Complex Intell Syst 4: 119–131.

[15] Arsenos, A., Kollias, D. and Kollias, S., 2022, June. A Large Imaging Database and Novel Deep Neural Architecture for Covid-19 Diagnosis. In 2022 IEEE 14th Image, Video, and Multidimensional Signal Processing Workshop (IVMSP) (pp. 1-5). IEEE.

[16] Turnbull, R., 2022. Cov3d: Detection of the presence and severity of COVID-19 from CT scans using 3D ResNets. arXiv preprint arXiv:2207.12218.

[17] Tan, W., Yao, Q. and Liu, J., 2022. Two-Stage COVID19 Classification Using BERT Features. arXiv preprint arXiv:2206.14861.

[18] Anwar, T., 2021. COVID19 Diagnosis using AutoML from 3D CT scans. In Proceedings of the IEEE/CVF International Conference on Computer Vision (pp. 503-507).

[19] https://keras.io/api/callbacks/reduce_lr_on_plateau/

[20] Ye, Andre, and Andre Ye. "A Deep Dive into Keras." Modern Deep Learning Design and Application Development: Versatile Tools to Solve Deep Learning Problems (2022): 1-48.

[21] Kim, Seongsoo, Hayden Wimmer, and Jongyeop Kim. "Analysis of Deep Learning Libraries: Keras, PyTorch, and MXnet." 2022 IEEE/ACIS 20th International Conference on Software Engineering Research, Management and Applications (SERA). IEEE, 2022.

[22] Kapoor, Amita, et al. Deep Learning with TensorFlow and Keras: Build and deploy supervised, unsupervised, deep, and reinforcement learning models. Packt Publishing Ltd, 2022.

[23] Kim, Changhyun, et al. "DESEM: Depthwise Separable Convolution-Based Multimodal Deep Learning for In-Game Action Anticipation." IEEE Access (2023).

[24] Das, Bijaylaxmi, Ayan Saha, and Sudipta Mukhopadhyay. "Rain Removal from a Single Image Using Refined Inception ResNet v2." Circuits, Systems, and Signal Processing 42.6 (2023): 3485-3508.

[25] Ouzar, Yassine, et al. "X-iPPGNet: A novel one stage deep learning architecture based on depthwise separable convolutions for video-based pulse rate estimation." Computers in Biology and Medicine 154 (2023): 106592.

[26] https://medium.com/@mygreatlearning/everything-you-need-to-know-about-vgg16-7315defb5918

[27] Singh, Arun, et al. "An Iris Recognition System Using CNN & VGG16 Technique." 2022 10th International Conference on Reliability, Infocom Technologies and Optimization (Trends and Future Directions)(ICRITO). IEEE, 2022.

[28] Aityan, Sergey K., and Sergey K. Aityan. "Confidence intervals." Business Research Methodology: Research Process and Methods (2022): 233-277.

[29] Takahashi, Kanae, et al. "Confidence interval for micro-averaged F 1 and macro-averaged F 1 scores." Applied Intelligence 52.5 (2022): 4961-4972.

[30] https://cpb-eu-w2.wpmucdn.com/blogs.lincoln.ac.uk/dist/c/6133/files/2022/07/mia_eccv_2022_leaderboard.pdf